\documentclass[aps,prb,twocolumn,superscriptaddress,showpacs]{revtex4-1}

\usepackage{amssymb,amsmath}
\usepackage[sort&compress]{natbib}
\usepackage{multirow}
\usepackage{xfrac}

\usepackage{graphicx}
\usepackage{dcolumn}
\usepackage{bm}
\usepackage{color}

\begin{document}

\title{Glass-like phonon scattering from a spontaneous nanostructure in AgSbTe$_2$}

\author{J. Ma$^{\star}$}
\affiliation{Quantum Condensed Matter Division, Oak Ridge National Laboratory, Oak Ridge, TN 37831, USA}

\author{O. Delaire$^{\star}$}
\email[Corresponding author: ]{delaireoa@ornl.gov}
\affiliation{Materials Science and Technology Division, Oak Ridge National Laboratory, Oak Ridge, TN 37831, USA}

\author{A. F. May}
\affiliation{Materials Science and Technology Division, Oak Ridge National Laboratory, Oak Ridge, TN 37831, USA}

\author{C. E. Carlton}
\affiliation{Department of Mechanical Engineering, Massachusetts Institute of Technology, Cambridge, MA 02139, USA}

\author{M. A. McGuire }
\affiliation{Materials Science and Technology Division, Oak Ridge National Laboratory, Oak Ridge, TN 37831, USA}

\author{L. H. VanBebber}
\affiliation{Department of Materials Science and Engineering, University of Tennessee, TN 37996, USA}

\author{D. L. Abernathy}
\affiliation{Quantum Condensed Matter Division, Oak Ridge National Laboratory, Oak Ridge, TN 37831, USA}

\author{G. Ehlers}
\affiliation{Quantum Condensed Matter Division, Oak Ridge National Laboratory, Oak Ridge, TN 37831, USA}

\author{Tao Hong}
\affiliation{Quantum Condensed Matter Division, Oak Ridge National Laboratory, Oak Ridge, TN 37831, USA}

\author{A. Huq}
\affiliation{Chemical and Engineering Materials Division, Oak Ridge National Laboratory, Oak Ridge, TN 37831, USA}

\author{Wei Tian}
\affiliation{Quantum Condensed Matter Division, Oak Ridge National Laboratory, Oak Ridge, TN 37831, USA}

\author{V. M. Keppens}
\affiliation{Department of Materials Science and Engineering, University of Tennessee, TN 37996, USA}

\author{Y. Shao-Horn}
\affiliation{Department of Mechanical Engineering, Massachusetts Institute of Technology, Cambridge, MA 02139, USA}

\author{B. C. Sales}
\affiliation{Materials Science and Technology Division, Oak Ridge National Laboratory, Oak Ridge, TN 37831, USA}


\begin{abstract}
Materials with very low thermal conductivity are of high interest for both thermoelectric and optical phase-change applications. Synthetic nanostructuring is most promising to  suppress thermal conductivity by scattering phonons, but challenges remain in producing bulk samples. We show that in crystalline AgSbTe$_2$, a spontaneously-forming nanostructure leads to a suppression of thermal conductivity to a glass-like level. Our mapping of the phonon mean-free-paths provides a novel bottom-up microscopic account of thermal conductivity, and also reveals intrinsic anisotropies associated with the nanostructure. Ground-state degeneracy in AgSbTe$_2$ leads to  the natural formation of nanoscale domains with different orderings on the cation sublattice, and correlated atomic displacements, which efficiently scatter phonons. This mechanism is general and points to a new avenue in nano-scale engineering of materials, to achieve low thermal conductivities for efficient thermoelectric converters and phase-change memory devices.
\end{abstract}

\maketitle

Nanostructured materials are among the strongest candidates for thermoelectric applications, as they offer a route to suppress thermal conductivity without hindering electrical properties \cite{Minnich-2009, Chen-2003}. The performance of thermoelectric materials is controlled by their figure-of-merit, $zT = \alpha^2 \sigma T/ \kappa_{\rm tot}$, where $\alpha$ is the Seebeck coefficient, $\sigma$ is the electrical conductivity, and $\kappa_{\rm tot}$ is the total thermal conductivity \cite{Wood, Goldsmid-book-2010, Snyder-2008}. Thus, high thermoelectric efficiency requires the lowest possible $\kappa_{\rm tot}$, while preserving a strong power factor, $ \alpha^2 \sigma$. Low thermal conductivity is also key in optical phase-change (PC) materials, which encode information in bits of either amorphous or crystalline state \cite{Wuttig-review}.  Because bit-state switching is triggered by melt-quenching, low thermal conductivity is important to decrease the heating power required to operate a PC memory chip \cite{Matsunaga-2011, Schneider-2010}. 

The semiconductor compound AgSbTe$_2$ is attractive for both thermoelectric and PC applications, because of its anomalously low, glass-like thermal conductivity  $\kappa_{\rm tot} \approx0.7\,{\rm Wm^{-1}K^{-1}}$ [\citenum{Hockings-1959, Wolfe-1960, Morelli-2008, Nielsen-2013, Zhang-2010, Detemple-2003}]. Its thermal conductivity is dominated by the lattice component, $\kappa_{\rm lat}$, corresponding to propagation of phonons \cite{Jovovic-2008}. The $\kappa_{\rm lat}$ in AgSbTe$_2$ is about three times lower than that of rock-salt PbTe at $300\,$K, and almost an order of magnitude lower at $100\,$K, as it is nearly independent of temperature. It was reported that $\kappa_{\rm lat}$ in AgSbTe$_2$ reaches the theoretical minimum, corresponding to phonon mean-free-paths limited to interatomic distances \cite{Wolfe-1960, Morelli-2008}. This glass-like $\kappa_{\rm lat}$ is key to achieve a high thermoelectric figure-of-merit, $zT \approx 1.3$ at 720$\,$K in pure form, and $zT_{\rm max} > 2$ when alloyed with PbTe in LAST \cite{Jovovic-2008, Hsu-Science-2004, CRC-handbook}. 
A detailed microscopic understanding of the origin of the anomalously low thermal conductivity in AgSbTe$_2$ has, however, remained elusive. Several factors have been pointed out, including strong anharmonicity \cite{Morelli-2008, Nielsen-2013}, and phonon scattering by differences in local force-constants of Ag$^{+}$ and Sb$^{3+}$ cations \cite{Ye-2008}. In addition, the phase-change character of AgSbTe$_2$ \cite{Detemple-2003} may reveal a relative instability originating from resonant bonding characteristics \cite{Wuttig-2008a, Wuttig-2008b}. Yet, no quantitative microscopic understanding of $\kappa_{\rm lat}$ has emerged, in part because phonon measurements were unavailable. 

Recent developments in inelastic neutron scattering (INS) enable a full mapping of the dynamic structure factor, $S({\bf Q}, E)$, for all phonon branches, across the entire Brillouin zone \cite{Delaire-PbTe-2011, Delaire-FeSi-2011, Pang-2013}. Here, we show how these developments offer an opportunity to derive a definitive microscopic account of the lattice thermal conductivity, thereby complementing more conventional experimental techniques. Our INS measurements of AgSbTe$_2$ provide critical insights into phonon scattering processes, from which we derive an unprecedented microscopic picture of thermal conductivity. Combining this novel approach with transmission electron microscopy (TEM), we demonstrate that this compound achieves glass-like thermal conductivity through the formation of a spontaneous nanostructure, which hinders the propagation of heat. The nanostructure in AgSbTe$_2$ originates from the ordering of cations into nano-scale domains with a correlation length $\sim3\,$nm. This spontaneous nanostructure efficiently scatters phonons, without the need for artificial nanostructuring. This mechanism should be of very high interest, as it points to an important direction for designing new materials with very-low thermal conductivity. 

\begin{figure*}
\centering 
\includegraphics[width=0.9\textwidth]{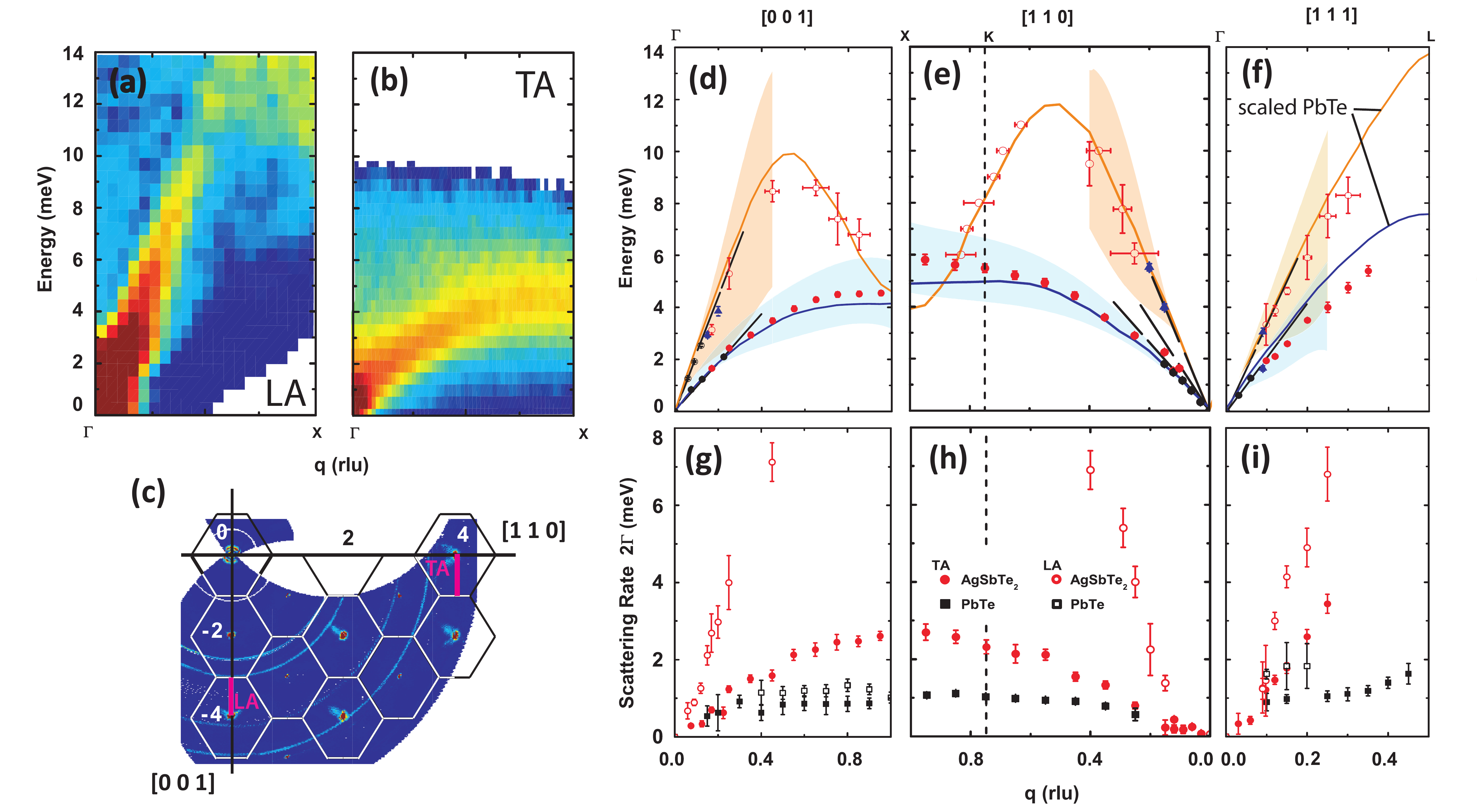}\\
\caption{Single crystal phonon data for $\delta$--phase AgSbTe$_2$ at $300\,$K. (a,b) Sample cuts of neutron scattering intensity $\chi \prime \prime ({\bf Q},E)=(1-\exp(-E/k_{\rm B}T))\,S({\bf Q}, E)$ measured with ARCS, for ${\bf Q}$ along red paths in (c), and showing broad LA and TA dispersions, respectively (red is higher intensity). (c) Reciprocal space map of elastic  scattering ($-1 \leqslant E \leqslant 1\,$meV) from ARCS measurement. (d,e,f) Dispersion curves and (g,h,i) phonon linewidths $2\Gamma=\tau^{-1}$, along high-symmetry directions, for acoustic phonons. The red dots/circles are measured with ARCS, black dots/circles with CTAX/CNCS, and the blue triangles with HB1A, respectively. Straight dashed lines correspond to sound velocities measured with RUS. Smooth dispersion curves are scaled PbTe data (see text). Phonon linewidths measured on PbTe at $300\,$K  are shown for comparison, and are considerably smaller than values for AgSbTe$_2$. Error bars on dispersions and linewidths result from statistical uncertainty in fits of phonon peaks (Supplementary).}
\label{Fig-dispersions}
\end{figure*}

Despite the apparent simplicity in chemical formulation, the crystal structure of  AgSbTe$_2$ shows a surprising degree of complexity. The 
$\delta$--phase of AgSbTe$_2$ forms slightly off stoichiometry in the pseudo-binary Ag$_2$Te -- Sb$_2$Te$_3$ phase diagram \cite{Maier-1963, Marin-1985,Sugar-Medlin-2009}, and one can express the composition in terms of the deviation from stoichiometry, $x$, as (Ag$_2$Te)$_{(1-x)/2}$(Sb$_2$Te$_3$)$_{(1+x)/2} =$ Ag$_{1-x}$Sb$_{1+x}$Te$_{2+x}$. The $\delta$--phase of AgSbTe$_2$ was long believed to crystallize in a cubic rock-salt structure with random Ag/Sb distribution on the cation sublattice \cite{Geller-1959}. However, a more recent single-crystal x-ray diffraction study proposed several possible ordered superstructures \cite{Quarez-JACS2005}, although these superstructures were not observed in subsequent TEM work \cite{Sugar-Medlin-2009, Sharma-2010, Sugar-Medlin-2011}. First-principles calculations of the ground-state structure also predicted several ordered superstructures with very close energies in both AgSbX$_2$ and AgBiX$_2$ (X$=$S, Se, Te) \cite{Nielsen-2013, Hoang-PRL2007, Barabash-PRL2008, Barabash-PRB2010}. In addition, it was shown with TEM that the ordering in AgBiS$_2$ and AgBiSe$_2$ tends to form nano-scale domains  \cite{Manolikas-1977}.  Our present results show that Ag and Sb in AgSbTe$_2$ also tend to order at the nanoscale, and that the resulting nanostructure efficiently scatters phonons.

Large single crystals of $\delta$--phase AgSbTe$_2$ were grown from melts with $x=0.1, 0.2$ (masses reaching $15\,$g), while stoichiometric crystals ($x=0$) were smaller ($1\,$g). Powder samples were synthesized for $x=0, 0.2$. Both neutron powder diffraction (NPD) and electron probe micro-analysis (EPMA) confirm that  samples with $x=0, 0.1$ contain a secondary Ag$_2$Te phase ($\sim$ 8$\%$ volume fraction for $x=0$).  No Ag$_2$Te was observed in the $x=0.2$ samples, on the other hand (with NPD, EPMA, and TEM). The EPMA compositions of the $\delta$--phase are Ag$_{22.2}$Sb$_{26.8}$Te$_{51}$, Ag$_{21.5}$Sb$_{27.2}$Te$_{51.2}$, and Ag$_{18.5}$Sb$_{29.3}$Te$_{52.2}$ for nominal $x=0, 0.1, 0.2$, respectively. Hereafter, we refer to all samples as AgSbTe$_2$, because these $\delta$--phase-dominated samples are found to behave similarly.

INS measurements were performed on single-crystals and powders, using both time-of-flight (ARCS \cite{ARCS} and CNCS \cite{CNCS}) and triple-axis (HB3, HB1A, CTAX) spectrometers at Oak Ridge National Laboratory. Details are given in the Supplementary Information (SI). Fig.~\ref{Fig-dispersions} summarizes our INS results on $\delta$--phase single-crystalline AgSbTe$_2$ (composition Ag$_{18.5}$Sb$_{29.3}$Te$_{52.2}$), concentrating on heat-carrying acoustic phonon dispersions along high-symmetry directions. The phonon linewidths, $2\Gamma_{j}(\bf{q})$, are inversely related to lifetimes, $\tau_{j}({\bf q})=(2\Gamma_{j}({\bf q}))^{-1}$. As may be seen in this figure, all acoustic modes are strikingly broad in energy, corresponding to very short phonon lifetimes. While the  transverse acoustic (TA) phonon branches and the lower part of the longitudinal acoustic (LA) branches are broad, but still well defined, the top of LA branches and the optical branches form a broad continuum of total width $\sim 10\,$meV. This precluded fitting peaks to extract optical dispersions, but it indicates that optical modes and LA modes of higher energy have only a marginal contribution to the lattice thermal conductivity (very short lifetimes). The phonon energies and linewidths were similar in the different ingots measured ($x=0, 0.1, 0.2$). 

\begin{table} [tph]
\caption{Sound velocities from resonant ultrasound measurements (RUS) and long-wavelength limit of INS dispersions, for  Ag$_{0.8}$Sb$_{1.2}$Te$_{2.2}$ at 300\,K. The values for $\kappa_{\rm lat}^{\rm calc.}$ correspond to the thermal conductivity contribution of each branch, assuming they are isotropic. Differences between directions illustrate anisotropies from both phonon group velocities and relaxation times.}
\begin{tabular}{cccc}
\hline
 Branch   &  $\quad v_g$ RUS (m/s)  &  $\quad v_g$ INS (m/s)   &  $\quad \kappa_{\rm lat}^{\rm calc.}$ Wm$^{-1}$K$^{-1}$  \\
\hline
TA&&\\
\hline
[001]  & 1,356 $\pm$ 14  &  1,407 $\pm$ 7 &  0.24 \\
\hline
[110]  & 1,356 $\pm$ 14  & 1,302 $\pm$ 8 &   0.53 \\
\hline
[111]  & 1,711 $\pm$ 18 & 1,746 $\pm$ 22 & 0.20 \\
\hline
LA &&\\
\hline
[001]  & 3,158 $\pm$ 160  & 3,073 $\pm$ 33 &  0.13 \\
\hline
[110]  & 2,879 $\pm$ 144  & 2,824 $\pm$ 55 &   0.21 \\
\hline
[111]  & 2,766 $\pm$ 138  & 2,687 $\pm$ 33 & 0.12 \\
\hline
\end{tabular}
\label{table1}
\end{table}

\begin{figure*} 
\includegraphics[width=0.8\textwidth]{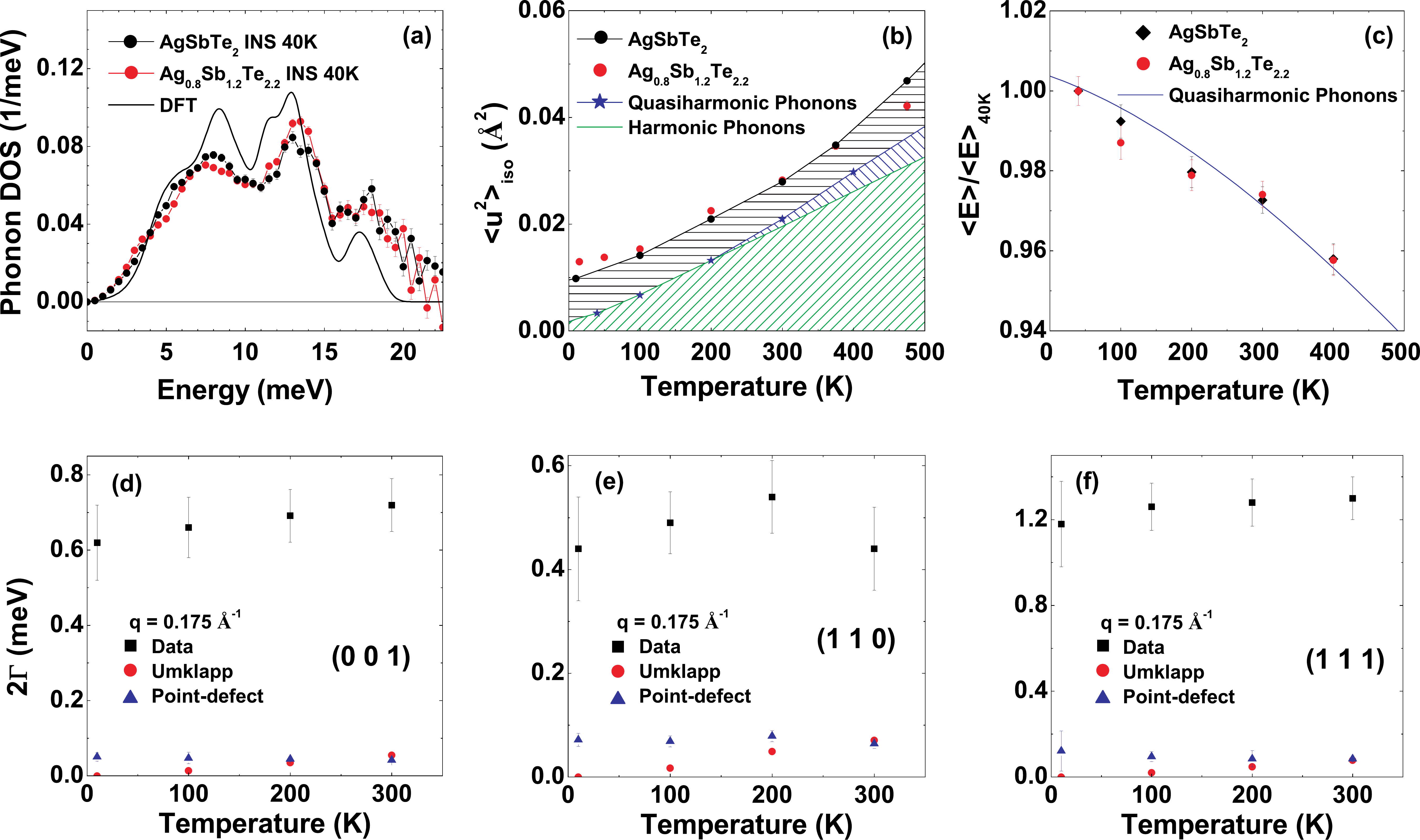} 
\caption{Temperature dependence of phonon properties, showing limited anharmonicity below $300\,$K. (a) Phonon DOS of powder samples Ag$_{1-x}$Sb$_{1+x}$Te$_{2+x}$, $x=0$ and $0.2$ at $300\,$K measured with ARCS ($E_i = 25\,$meV), and calculated from first-principles for a $R\bar{3}m$ superstructure, in the harmonic approximation. (b) Mean-square-displacement $\langle u^2 \rangle$ of AgSbTe$_{2}$ sample (black dots and line), and Ag$_{0.8}$Sb$_{1.2}$Te$_{2.2}$ sample (red dots) measured with NPD, and calculated from the phonon DOS of AgSbTe$_{2}$ in (a). Blue stars are quasiharmonic $\langle u^2 \rangle$ from DOS data at respective temperatures, and the green line is the harmonic $\langle u^2 \rangle$ from the DOS measured at 40K. This comparison reveals a static component of atomic displacements besides the phonon contribution. (c) Relative change in average phonon energy $\langle E(T) \rangle / \langle E \rangle_{\rm 40K}$ \textit{vs} $T$ for AgSbTe$_{2}$ (black diamonds) and Ag$_{0.8}$Sb$_{1.2}$Te$_{2.2}$ (red dots), compared with quasi-harmonic model for AgSbTe$_{2}$, based on the measured lattice parameter and an average Gr\"uneisen parameter $\bar{\gamma}=2.05$ (from Ref. \citenum{Morelli-2008}), showing good agreement ($\langle E \rangle$ values and error bars are derived from the measured DOS, Fig.~S5). (d,e,f) Measured linewidths of TA phonon modes with $q=0.175\,$\AA$^{-1}$ along high-symmetry directions, as a function of temperature, and calculations of contributions from umklapp scattering, and from point-defect scattering (error bars are propagated from data in Fig.~\ref{Fig-dispersions})}.
\label{T_dependence}
\end{figure*}

The slopes of the phonon dispersions for long wavelengths (low $q$) are in excellent agreement with our resonant ultrasound (RUS)  measurements of sound velocities (dashed straight lines in Fig.~\ref{Fig-dispersions}-d,e,f and Table~\ref{table1}). Our results for sound velocities are considerably lower than the polycrystalline estimate reported in the literature \cite{Wolfe-1960}. The smooth dispersion curves in Fig.~\ref{Fig-dispersions}-d,e,f were obtained from the acoustic phonon dispersions of PbTe, with energies scaled by the ratio of average masses, $(M_{{\rm AgSbTe}_2}/M_{\rm PbTe})^{1/2}$. The good agreement between the scaled PbTe phonon dispersions and the dispersions of AgSbTe$_2$ indicates that the average force-constants in both materials are comparable overall, although phonon dispersions of AgSbTe$_2$ are relatively softer along the $[111]$ direction.

The linewidths (FWHM) of acoustic phonons, $2 \Gamma_{{\bf q}, j}$, obtained after correction for the instrumental resolution (details in SI),  are compared to those of single-crystalline PbTe \cite{Delaire-PbTe-2011} in Fig.~\ref{Fig-dispersions}-g,h,i. The transverse acoustic phonons in PbTe have much narrower linewidths than in AgSbTe$_2$. This was confirmed with measurements on both ARCS and CNCS, which gave very similar phonon linewidths, $2\Gamma$ (SI). This is consistent with $\kappa_{lat}$ being three times larger in PbTe than in AgSbTe$_2$ at $300\,$K, despite lower group velocities in PbTe. We also note that the TA linewidths in AgSbTe$_2$ are especially broad along $[111]$.

In Fig.~\ref{T_dependence}, we present results for the temperature dependence of phonon energies and linewidths in AgSbTe$_2$ ($x=0.2$). Our results show that, while some anharmonicity is present, it is not the dominant source of phonon scattering. In particular, as may be seen in Fig.~\ref{T_dependence}-d,e,f, the TA phonon linewidths show no temperature dependence between $10\,$K and $300\,$K, within experimental uncertainty. This indicates that the dominant scattering mechanism for TA phonon scattering is independent of temperature, ruling out anharmonicity. Rather, the nanostructure is the main source of phonon scattering in this regime, although anharmonicity could become more significant for $T>300\,$K.

The phonon density of states (DOS) measured with INS on powders ($x=0$ and $0.2$) is shown in Fig.~\ref{T_dependence}-a. Both samples have the same DOS within experimental uncertainties, indicating that the Ag$_2$Te impurity phase for $x=0$ only has a small effect (as does the difference in $\delta$--phase composition). The DOS exhibits  three broad peaks centered around 7, 13, and 17\,meV,  and phonon cutoff of about 20\,meV, in good agreement with our first-principles calculations (see below). Because the masses and cross-sections are similar for Ag, Sb, and Te, the spectrum obtained from INS does not significantly distort the phonon DOS. The average phonon energy ratio, compared with PbTe, $\langle E_{{\rm AgSbTe}_2} \rangle / \langle E_{\rm PbTe} \rangle = 1.20$ ($300\,$K), is very close to the inverse square-root of the mass ratio $(M_{{\rm AgSbTe}_2} / M_{\rm PbTe})^{1/2}=1.17$, again indicating that the underlying force-constants are comparable, on average, in the two materials. 

We observe a softening of the phonon DOS from 40 to 400$\,$K (SI), which is an expected effect of moderate anharmonicity. The amount of softening is well accounted for by thermal expansion, however, thus being compatible with the quasiharmonic (QH) model.
Figure~\ref{T_dependence}-c shows the relative change in $\langle E \rangle$ in both powder samples, as a function of temperature, as well as the expected QH behavior derived from the NPD measurements, and the Gr\"{u}neisen parameter from Ref.~\citenum{Morelli-2008}. As can be seen on this figure, the temperature dependence of phonon energies in the powder samples follows the QH model quite well up to 400$\,$K. The QH model also provided a very good account for the heat capacity measured up to $300\,$K (see SI). The optical peaks in the phonon DOS showed some broadening with increasing $T$ (SI), which may indicate more anharmonicity of the optic modes, possibly because of overlap with the top of LA branches. 

In Fig.~\ref{T_dependence}-b, we compare the temperature dependence of the isotropic average thermal displacement measured with NPD, $\langle u^2 (T) \rangle_{\rm NPD}$, to the vibrational component derived from the measured phonon DOS. The $\langle u^2 \rangle$ component due to harmonic vibrations, $\langle u^2 \rangle_{\rm har}$, is systematically lower than the total atomic displacement, with a constant offset, $\langle u^2 \rangle_{\rm NPD}-\langle u^2 \rangle_{\rm har} \sim 0.008\,$\AA$^2$ for $T \leqslant  400\,$K. The softening of the phonon DOS with increasing $T$, resulting in the quasiharmonic value $\langle u^2 \rangle_{\rm qh}$, does not account for this effect. Rather, the offset reveals static displacements, which are a temperature-independent source of phonon scattering. These static displacements originate from atomic relaxations in and between ordered nanodomains.

\begin{figure*}
\includegraphics[width=0.95\textwidth]{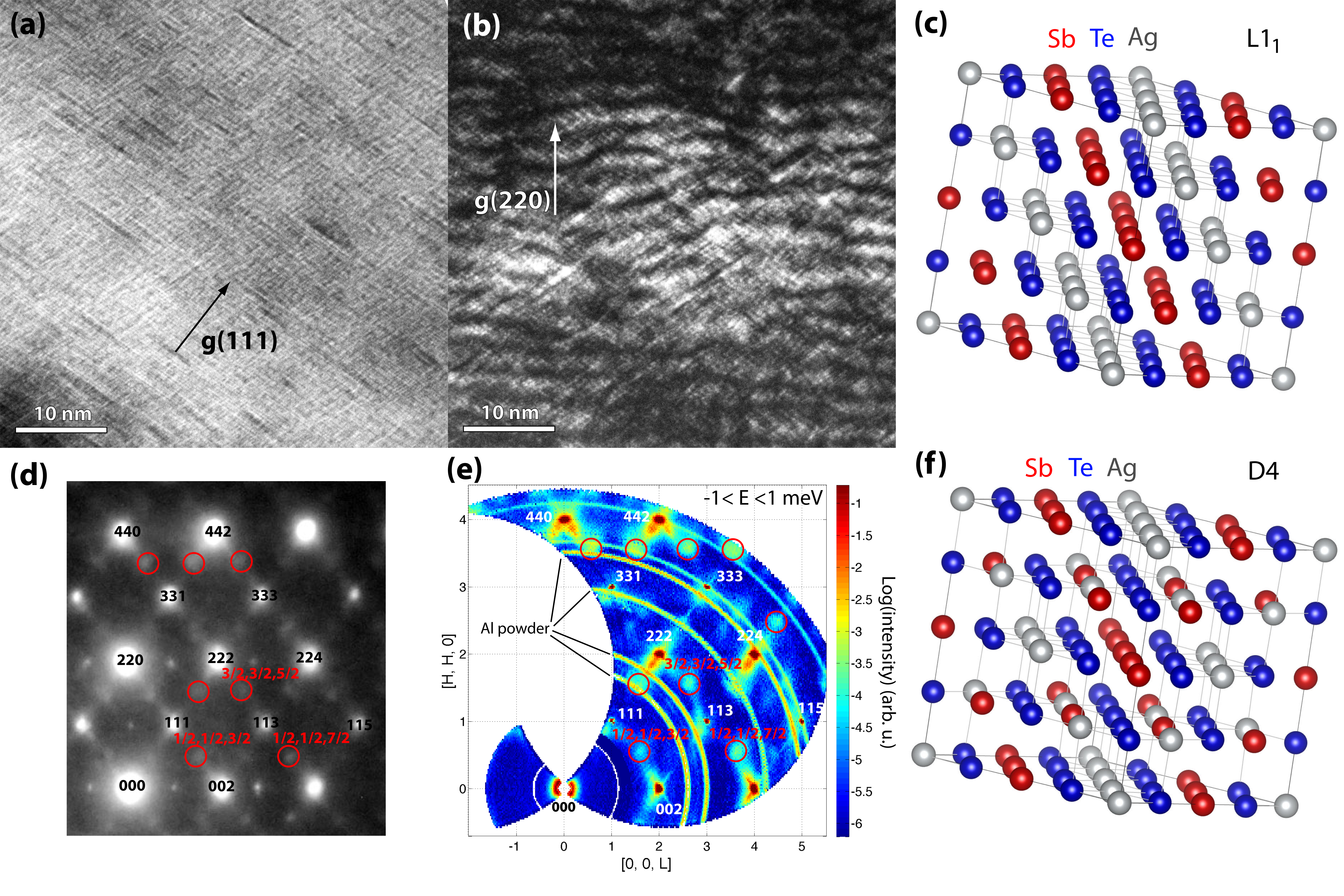}\\
 \caption{Transmission electron microscopy (TEM) and neutron diffuse scattering, revealing the nanostructure in $\delta$--phase AgSbTe$_{2}$. (a) 2-beam (111) bright-field diffraction contrast TEM showing the superlattice nanodomains and associated strains. (b) 2-beam (220) dark-field diffraction contrast showing diagonal tweed nanodomains and oscillating strains. (d,e) Electron diffraction pattern and elastic neutron scattering ($-1 \leqslant E \leqslant 1\,$meV), indexed to the cubic rock-salt parent structure. The pink circles show superstructure ordering spots at $\frac{1}{2} (h,k,l)$ with $h,k,l$ odd ($L$ points of Brillouin zone). Extra spots along ${\bf g}(002)$ and ${\bf g}(220)$ in (d) are artifacts. In panel (e), the lobes around rock-salt diffraction spots, extending along $\langle 111 \rangle$ directions, reveal correlated local atomic displacements. (c,f) Competing structures for ordering of Ag and Sb cations. }
\label{nanostructure}
\end{figure*}

We performed first-principles density functional theory (DFT) calculations on ordered superstructures of AgSbTe$_2$, including the $D4$ and $L1_1$ cation superstructures (Fig.~3-c,f) previously predicted as candidate ground-states (details in SI) \cite{Hoang-PRL2007, Barabash-PRL2008, Barabash-PRB2010}. Upon relaxing ion positions, we find systematic displacements of Te toward Ag cations in $\langle 100 \rangle$ first nearest-neighbor (1NN) bonds: $d_{\rm Ag-Te}=3.03\,{\rm \AA}$ \textit{vs} $d_{\rm Te-Sb}=3.11\,{\rm \AA}$. These static displacements likely contribute to phonon scattering, as mentioned above. Our DFT phonon calculations on the ordered structures predict a significant overlap of optic and longitudinal acoustic modes, although some of this effect is associated with the band folding induced by the use of a supercell. The effect is likely intrinsic, however, based on the fact that all atomic masses are similar in this material, which does not promote the same energy-separation of acoustic and optic modes as in PbTe, where acoustic modes (in particular TA modes) mainly involve Pb motions, and optical modes mainly involve Te motions. This is corroborated by the good agreement between the measured phonon DOS and our calculation for a $R\bar{3}m$ ordered superstructure (64-atom cell) (Fig.~\ref{T_dependence}-a). The overlap of acoustic and optic modes in AgSbTe$_2$ could lead to enhanced interaction between the high-frequency acoustic modes and low-frequency optic modes. The results of our phonon calculations are in overall agreement with previous studies \cite{Hoang-PRL2007, Barabash-PRL2008}.

As previously mentioned, the lack of $T$-dependence of phonon relaxation times, $\tau_{{\bf q},j}=1/(2\Gamma_{{\bf q},j})$, indicates that anharmonic phonon-phonon scattering is not the dominant phonon scattering mechanism, at least up to room temperature, for heat-carrying TA phonons. As originally shown by Peierls, phonon-phonon umklapp scattering follows a $T$ dependence scaled by the Debye temperature, $\theta_D$  [\citenum{Peierls-1929}]. From the phonon DOS for Ag$_{0.8}$Sb$_{1.2}$Te$_{2.2}$ at $300\,$K, we obtain $\theta_D=4/3 \langle E \rangle =155\,$K (and $153\,$K for the AgSbTe$_{2}$). Thus phonon-phonon scattering would be expected to give a strong $T$-dependence of the scattering rates in the measured range of temperature, which is not observed. We calculated the expected phonon linewidths from umklapp processes, using the formula derived by Slack and Galginaitis \cite{Slack-1964} (details in SI). The results for linewidths $2\Gamma_{\rm u} = \tau_{\rm u}^{-1}$ for $q=0.175\,{\rm \AA}$ (close to  the dominant contribution to $\kappa_{\rm lat}$) are plotted in Fig.~\ref{T_dependence}. As can be seen, $2\Gamma_{\rm u}$ is much smaller than the measured linewidths, at most 15\% of the observed $2\Gamma$ at $300\,$K. 

Point-defects constitute another possible source of phonon scattering, owing to the mixed nature of the Ag/Sb cation sublattice, and possible vacancies \cite{Barabash-PRL2008}. The results for $2\Gamma_{p.d.}=\tau_{p.d.}^{-1}$ calculated with the model of  Krumhansl and Matthew \cite{Krumhansl} are plotted in Fig.~\ref{T_dependence}, taking into account both the force-constant difference of Ag and Sb sites\cite{Ye-2008}, and the additional effect of 5\% vacancies. Point-defect scattering accounts for about 10--15\% of the measured linewidth. From the analysis of phonon lifetimes for typical scattering sources, one must conclude that a more significant defect structure is responsible for the large, $T$--independent phonon linewidths. Such a conclusion is directly supported by our observations with TEM and diffuse neutron scattering data.


TEM observations reveal that all $\delta$--phase samples (Ag$_{1-x}$Sb$_{1+x}$Te$_{2+x}$, $x=0, 0.1, 0.2$) exhibit a pervasive nanostructure, which includes cation-ordering in nano-domains, and correlated atomic displacements resulting in strains. Superlattice reflections are observed half-way between all-even and all-odd diffraction spots of the parent rock-salt structure, as highlighted by red circles in Fig.~\ref{nanostructure}-d,e. They correspond to the $L$ points of the FCC Brillouin zone ($\frac{1}{2} (h,k,l)$ with $h,k,l$ odd in rock-salt reciprocal lattice). This reveals a doubling of the lattice period along $\langle 111 \rangle$ directions in real space. This is compatible with a segregation of Ag and Sb ions to alternating $(111)$ crystallographic planes, as expected for the proposed $L1_1$ ground-state superstructure \cite{Hoang-PRL2007, Barabash-PRL2008, Barabash-PRB2010}. It is also compatible with the  related ordering in the $D4$ structure of nearly degenerate energy \cite{Hoang-PRL2007, Barabash-PRL2008, Barabash-PRB2010}. These two structures are illustrated in Fig.~\ref{nanostructure}-c,f. Thus, it appears that energy-degeneracy between competing cation orderings leads to the formation of a complex ordering pattern, with multiple ordering nanodomains. Because the 4 variants of $L1_1$ (corresponding to 4 possible $\langle 111 \rangle$ directions) and the D4 variant have nearly the same energy, the material is in a ``frustrated'' state, and remains ordered only locally. The width of the superlattice peaks in single-crystal neutron data provides a correlation length for ordered domains $\xi \sim 3\,$nm. In addition, slight offsets of the superstructure peaks from $L$ points were observed, indicating possible further modulations of ordered nano-domains.  Thus, we find a similar behavior as previously reported in  AgBiSe$_2$ and AgBiS$_2$ \cite{Manolikas-1977}, although the nanostructure appears somewhat more diffuse in Ag$_{1-x}$Sb$_{1+x}$Te$_{2+x}$. It is plausible that similar orderings and nanostructures could occur in related ABX$_2$ compounds.

The nanostructure was imaged with 2-beam condition in the TEM. In Fig~\ref{nanostructure}-a, elongated features a few nanometers across are observed in the $\langle 111 \rangle$ planes, associated with strain contrast of nanodomains. Their shape could not be uniquely determined, but they are either platelets or needles. In Fig~\ref{nanostructure}-b, an oscillating contrast is clearly seen, forming light and dark ripples of $3\,$nm period, associated with strain contrast. 
Our DFT simulations for the $L1_1$--ordered structure predict a tendency toward rhombohedral distortion (about $0.8^{\circ}$ flattening of cuboid along rhombohedral axis). Thus, cation ordering in nanodomains of $L1_1$ variants should tend to strain the lattice along different $\langle 111 \rangle$ directions, while $D4$ nanodomains would not. While no rhombohedral splitting of peaks was observed in our NPD diffraction patterns, it may be because constraints between neighboring nanodomains suppress the distortion. We note that the strain contrast in Fig~\ref{nanostructure}-b does not directly superimpose with the contrast in Fig~\ref{nanostructure}-a, although its lateral periodicity is in good agreement with the ordering correlation length ($\sim 3\,$nm). In addition, a modest density of stacking faults was also observed, likely associated to the excess Te in the $\delta$--phase and the formation of Te-Te double-layers, in agreement with prior observations \cite{Sharma-2010, Sugar-Medlin-2011}.

Further, HRTEM and fast-Fourier transforms (FFT) support the superlattice interpretation of the diffraction patterns (SI). The HRTEM images clearly show layers parallel to $(111)$ in agreement with the Ag-Te-Sb-Te-Ag stacking sequence (periodicity $\sim 7\,$\AA) of (111) planes in  the $L1_1$ structure. The FFT's obtained from different regions of the HRTEM reveal different $L1_1$ ordering directions, while other regions appeared compatible with $D4$ order, confirming the local nature of the ordering.


\begin{figure}
\includegraphics[width=0.45\textwidth]{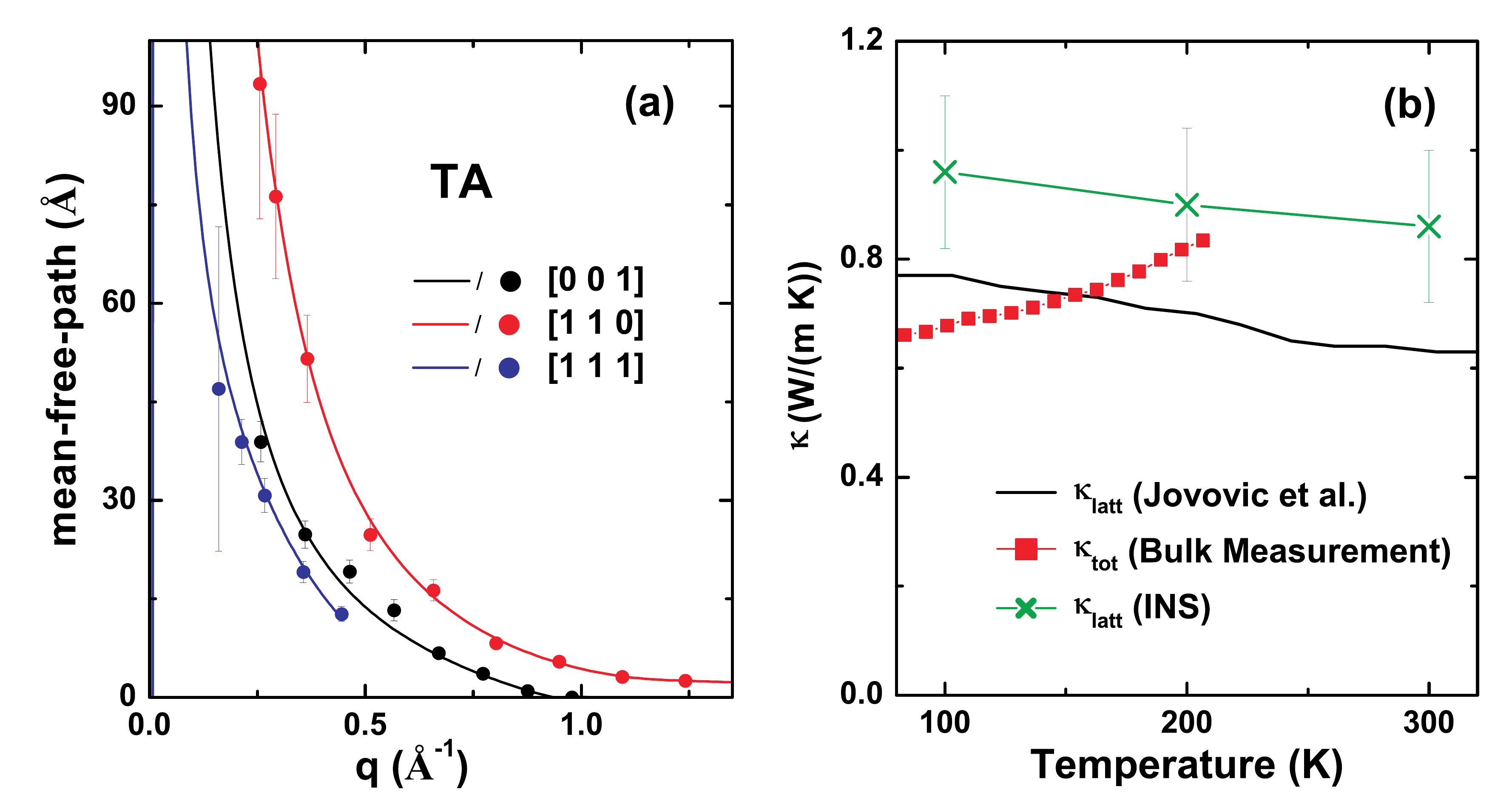}\\
\caption{Mean-free-paths and lattice thermal conductivity derived from INS data. (a) Mean-free-paths of TA phonons in Ag$_{0.8}$Sb$_{1.2}$Te$_{2.2}$ along different directions, derived from values for $\tau$ and dispersion group velocities in Fig.~\ref{Fig-dispersions}, and their associated uncertainties. (b) Thermal conductivity from bulk transport measurements, and calculated from measured mean-free-paths (details in  Supplementary).} 
 \label{figure4}  
\end{figure}

The length-scale of the observed nanostructure is in good agreement with the measured phonon mean-free-paths, $\Lambda_j({\bf q})$. In addition, we observe an anisotropy in the lifetimes and mean-free-paths, which could be related to the anisotropy of the nanostructure. The dominant mean-free-path of TA phonons, $\Lambda_{\rm TA, dom}$, was estimated by integrating the data in Fig.~\ref{figure4}, weighted by their normalized contribution to the thermal conductivity. The results are $\Lambda_{\rm TA, dom}=32, 45, 65\,{\rm \AA}$ along [111], [100], and [110], respectively. The TA mean-free-paths are thus shortest along $[111]$, despite the larger group velocity in that direction. In the diffusive regime,  the total lattice thermal conductivity  $\kappa_{\rm lat}$ must be isotropic for a system of average cubic symmetry, as $\kappa_{\rm lat}$ is a rank-2 tensor in that limit. However, in the ballistic regime, $\kappa_{\rm lat}$ is governed by the rank-4 elastic tensor (group velocities), as well as possible anisotropies in the microscopic phonon scattering mechanisms themselves \cite{Wolfe-book}. Because the mean-free-paths are so short in AgSbTe$_2$, macroscopic samples are always in the diffusive regime, even at low temperatures. Our  thermal conductivity measurements in oriented single-crystals confirm this (see SI).  We point out that our values of $\Lambda_{\rm TA, dom}$ are an order of magnitude larger than interatomic distances, but $\Lambda$ for optical modes could be shorter. The magnitude of mean-free-paths for heat-carrying TA phonons confirms the importance of the nanostructure in scattering phonons.

The contributions of phonon dispersions along different directions were averaged to obtain the total, isotropic $\kappa_{\rm lat}$ in the diffusive limit. The total thermal conductivity was calculated by summing individual phonon contributions, with an isotropic average (details in SI). The effective, isotropic contributions of different phonon branches to $\kappa_{\rm lat}$ are listed in Table~\ref{table1}.  Overall, the TA modes contribute more than $2/3$ of the total $\kappa_{\rm lat}$, owing to the limited LA contribution at high-$q$, where they overlap with optical modes. The estimate of $\kappa_{\rm lat}$ based on the measured phonon mean-free-paths is in good agreement with the values obtained from bulk transport measurements, as shown in Fig.~\ref{figure4}. This result establishes neutron scattering as a powerful tool to probe the microscopic origins of lattice thermal conductivity.

Our neutron scattering and TEM results show that a pervasive nanostructure develops in $\delta$--phase AgSbTe$_{2}$, similar to that in AgBiSe$_2$ and AgBiS$_2$ \cite{Manolikas-1977}. This spontaneous nanostructure involves both local ordering of Ag and Sb on the cation sublattice, and correlated atomic displacements, that combine to scatter phonons on length scales of a few nanometers. The nanostructure may act in concert with anharmonic bonding in suppressing $\kappa_{\rm lat}$, especially above $300\,$K. We note that variations in thermal conductivity of AgBiSe$_2$ upon heat treatments \cite{Nielsen-2013} may involve changes in short-range ordering and size of nanodomains \cite{Manolikas-1977}. 
Materials spontaneously forming nanoscale inhomogeneities alleviate the need for artificial nanostructuring, and are therefore of great interest \cite{Minnich-2009, Biswas-2011, Biswas-2012, Pei-2011}.  Compared with other uses of nanostructuring, the phonon scattering mechanism discussed here does not require alloying, which may be advantageous in decoupling the optimization of thermal and electronic transport.  Ordered nano-scale domains could occur in other materials where mixed species  share a sublattice, and first-principles simulations could be used to systematically search for near-degeneracies in ordering energies. Our findings could thus provide guidelines for finding other materials with low thermal conductivities for thermoelectric and phase-change applications.


\begin{center}
Acknowledgements
\end{center}
Simulations and integration of results (O.D.), as well as synthesis and characterization (A.F.M., M.A.M., B.C.S) were supported by the US Department of Energy, Office of Basic Energy Sciences, Materials Sciences and Engineering Division. Neutron scattering (J. M.) and electron microscopy (C. E. C., Y. S.-H.) were supported by the US Department of Energy, Office of Basic Energy Sciences, through the S3TEC Energy Frontier Research Center, DESC0001299. L.H.V. and V.M.K. acknowledge support provided by the the Joint Directed Research and Development program of the UTK Science Alliance. The Oak Ridge National Laboratory's Spallation Neutron Source and High-Flux Isotope Reactor are sponsored by the Scientific User Facilities Division, Office of Basic Energy Sciences, US Department of Energy. 

\begin{center}
Author Contributions
\end{center}
${}^\star$ Both first-authors contributed equally to the study. O.D. designed the study. O.D. and J.M. performed and analyzed neutron scattering experiments, and wrote the manuscript. A.F.M., M.A.M. and B.C.S synthesized and characterized the samples. C.E.C and Y. S.-H. performed and analyzed electron microscopy. L.H.V. and V.M.K. performed and analyzed ultrasound measurements. O.D. performed first-principles simulations. D.L.A, G.E., T.H., A.H. and W.T. helped with neutron data acquisition.

\begin{center}
Additional Information
\end{center}
The authors declare no competing financial interests.


\bibliographystyle{apsrev}

\newpage


\end{document}